\documentstyle[epsf,epsfig,12pt]{article}
\begin{document}

\vspace{-20mm}
\begin{flushright} 
{\bf LAL 03-08}\\
March 2003\\
\end{flushright}

\vspace{0.5 cm}
\Large
\centerline {\bf Present and future sensitivity to a Z-prime}
\normalsize
\vskip 0.2  cm
\centerline {F. RICHARD}
\smallskip
\centerline {\bf Laboratoire de l'Acc\'el\'erateur Lin\'eaire,}
\centerline {\it  IN2P3-CNRS et Universit\'e de Paris-Sud,  BP 34 - F-91898 Orsay Cedex, France}
\noindent
\vskip 0.2 cm
\begin{abstract}
Various extensions of the Standard Model, i.e. string inspired groups like E$_6$, extra dimensions, 'deconstruction', predict additional vector bosons. In most cases these bosons can mix with the SM Z and this translates into a very high sensitivity of precision measurements which can be performed at the Z resonance. This
aspect will be illustrated using SLD/LEP data and the possible impact on the Higgs boson mass indirect determination. An extrapolation to the accuracies which could be reached at FLC will be discussed.
With accuracies available at GigaZ it will become possible to elucidate the origin the Z' and the symmetry breaking mechanism responsible for the mixing. Two examples show the possible impact of FLC on the LHC program.    
\end{abstract}
\vskip 0.5 cm
\centerline{\it Work presented at the LHC/LC Study Group (CERN July 5 2002 and February 14 2003)} 
\vskip 0.5  cm
\section{Introduction}

Unification of forces within SUSY is an essential paradigm resulting from LEP1/SLD precision measurements. 
The simplest GUT group is SU(5) but extended symmetries\cite{E6} like E$_6$ or its subgroup SO(10) provide an additional heavy Majorana neutrino which could generate neutrino masses through the see-saw mechanism. As pointed out in \cite{ham} E$_6$, a group which naturally occurs within string theory, is a good candidate since it allows symmetry breaking in steps and therefore the coexistence of a TeV scale Z' together with a GUT scale Majorana neutrino. \par
Similarly, left-right symmetry restoration at the TeV scale, which may also be embedded in SO(10),
would imply the existence of a Z$_{LR}$ vector boson.  \par
Extra dimensions constitute an alternate to the SUSY extension of the 4D Poincar\'e group. These theories can solve the hierarchy problem by lowering the 
effective Planck scale\cite{ND} down to a TeV. If this scenario is realized in a string theory of quantum gravity, one should observe string excited states of the SM particles also a the TeV scale\cite{string}. 
String models naturally predict extra Z' with masses which could be well below the string scale. This is illustrated in $\cite{Iban}$, providing an example of a fairly complex Z' sector. One of these Z' could be ten times lighter than the string scale, eventually providing the only visible indication of this structure.   \par
In models with TeV$^{-1}$ extra dimensions, new heavy bosons are predicted, such as in \cite{Anton}, in which one expects Kaluza Klein excitations of 
$\gamma$/Z with large couplings to fermions. In contrast, in the Universal Extra Dimension scheme, where all SM have KK partners, a conservation law forbids such couplings but, as we shall see, the new scale still 
manifest its presence in LEP/SLD precision measurements.\par 
'Deconstruction', inspired by extra dimensions, is proposed as an alternate to SUSY and higher dimensions\cite{Arkani} to provide a solution to the hierarchy problem. In this scenario, also known as 'Little Higgs', 
new symmetries imply the existence of a rich spectrum of new particles, in particular light gauge bosons. The
Higgs boson appears as a pseudo-Goldstone boson and is protected by the global symmetry from 1-loop quadratic
divergences.   \par  
\newpage
Finally, the BESS model appears as an other dynamical electroweak symmetry breaking and also predicts the existence of new gauge bosons\cite{casal}. \par
This paper will analyse the influence of these various models on LEP/SLD precision measurement (PM) and in 
particular the impact they could have on the indirect Higgs boson mass determination with the basic question:
{\it can the effect of a Z' on PM hide the presence of a heavy Higgs boson ?} \par
I will also discuss the possible consequences for future colliders. I will argue that  
in the presence of so many competing models it might not be easy to interpret a Z'-like signal found at LHC. It may even occur that a Z'-like signal would be the only 'smoking gun' which would herald the presence of an extension of the SM, hence the importance of preparing the proper tools to cope with this scenario. I will therefore recall how a future linear collider, FLC, can provide specific inputs to this complex phenomenology.   \par

4 experimental methods are used to search for Z' imprints: \par
- direct searches through the leptonic channel at hadronic colliders \par
- semi-direct searches at LEP2 and FLC through interference between Z  

and Z' amplitudes giving  access to the vector and axial couplings given the Z' mass \par
- indirect searches through atomic parity violation APV\cite{atom}\par
- indirect searches through precision measurements at the Z resonance in e$^+$e$^-$. \par

\begin{table}[t]
\centering
\caption{Limits on Z' masses in GeV}
\vskip 0.5 truecm
\begin{tabular}{|c|c|c|c|}
\hline Model &$\chi$ & $\psi $ & $L-R$  \\  
\hline
\hline LEP2 & 630 & 510 & 950  \\
\hline FNAL & 595 & 590 & 630  \\
\hline Atomic Parity &   730  &  - & 790\\
\hline
\end{tabular}
\end{table}

Present limits based on the first 3 methods are recalled in table 1. 
In the future, the discovery range will reach about 5 TeV at LHC. FLC can fully cover this range an determine the axial and vector couplings of a Z'\cite{tesla}.\par
The last method will be the main emphasis of this paper. I will argue that in most models there is mixing between Z and Z' and that this mixing affects considerably precision measurements at the resonance, therefore providing an additional and very sensitive tool to understand the origin of this particle. 
I will also extrapolate this analysis to the LHC/FLC era with emphasis on the so called GigaZ mode of operation of FLC which improves dramatically\cite{tesla} on LEP/SLD. \par
Finally, possible impact of FLC results on LHC operations will be discussed.
\section{LEP/SLD results}

Results from LEP/SLD are consistent with the minimal SUSY extension of SM, called MSSM, which predicts a Higgs boson with a mass 115-135 GeV and, in most scenarios, no other indirect contribution from the SUSY partners. \par

For some time, there has been concern about the discrepancy\cite{PM} between the most precise measurements giving sin$^2\theta^{lept}_{eff}$ - the left-right asymmetry measured at SLD and the forward-backward asymmetry for b quarks measured at LEP1- which reaches 3 s.d. Interpreted in terms of a Higgs boson mass, the latter corresponds to a 600 GeV Higgs boson, while the former is consistent with a Higgs boson of 40 GeV. Recent results\cite{PM} on the W mass have added
to this concern since they also correspond to a very light Higgs boson. One should 
however not forget that these results
are correlated through the uncertainty on the top mass and that the discrepancy with a 114 GeV Higgs boson, the present lower limit on the SM Higgs boson given by LEP2, is only at the 2 s.d. level.\par
The b quark asymmetry discrepancy has not received any simple theoretical or experimental interpretation. There could be heavier fermions in extended symmetries, like E$_6$, which mix with b quarks, hence the observed effect. It turns out that life is more complicated since there is no discrepancy on the cross-section, which is hard to accommodate in E$_6$. Other ideas have been suggested, like more exotic groups\cite{Ma} or mirror 
fermions\cite{Carlos}, which could accommodate both features. \par
One cannot exclude an experimental bias for this difficult measurement which requires not only to tag the presence of b quarks but also to measure the jet charges. For good reasons, experts defend this result on the basis that diverse methods are used in these analyses and that all results appear to be in good agreement. \par

Whichever of these explanations is true, in the framework of Z' models with universal couplings to fermions, 
one is entitled to attempt an interpretation of the data ignoring the b sector. This has already been done assuming that there could be very light SUSY particles\cite{Altfb}, while here I will assume that SUSY particles do not influence LEP/SLD precision measurements (PM) and try the effect of a Z'. \par

Three LEP/SLD observables will be used in the present analysis: \par
- the W mass\par
- sin$^2\theta^{lept}_{eff}$ as determined from the leptonic asymmetries, primarily at SLD \par
- the leptonic width $\Gamma_l$ \par
These quantities, as shown explicitly in the appendix, are complementary in the sense that they are differently influenced by Z' parameters: \par
- the W mass is independent of the couplings and is always increasing with the Z' contribution\par

- sin$^2\theta^{lept}_{eff}$ depends primarily on the vector part of Z' couplings \par
- $\Gamma_l$ depends primarily on the axial part.\par

Above features are not strictly true in the 'Little Higgs' scenario for which there can
be contributions from B' and W'$_3$ components but, unless there is no B'-Z mixing (for x=1, as discussed in the appendix) the major contribution comes from B' which tends to be the lightest Z'. \par
In figures 1,2 and 3 are represented the measured values with their errors
and the SM prediction assuming a Higgs boson mass of 115 GeV and taking into account errors from
the top mass and $\alpha(M_Z)$ taken from \cite{alph}.
With present accuracies, one cannot claim any significant disagreement but a preference for a Higgs boson lighter than 114 GeV, already excluded by direct searches. \par

 On the same figures one can see what can be expected with FLC accuracies. The top mass and the Higgs boson mass will be known to better than 100 MeV, while the reduction in error on  $\alpha(M_Z)$ by a factor 3, anticipated in\cite{tesla}, is a guess. With these figures, the SM ellipse would shrink into a point, while the errors on the W mass and sin$^2\theta^{lept}_{eff}$ are greatly reduced.    
\section{Implications of Z-Z' mixing at LEP/SLD }
As pointed in reference\cite{Peswel}, the presence of a Z' with a TeV mass can fake a light Higgs boson, as indicated by the LEP1/SLD PM. This is shown in 
figures 1, 2 and 3 where the various Z' 
predictions appear as straight lines: at infinite mass one recovers the SM point while a lighter Z' always corresponds to a heavier W.
\subsection{E$_6$}
In E$_6$, the Z' properties depend on the mixing of the two U(1) components called $\psi$ and $\chi$ with a mixing angle which is called $\theta$:
\newpage
$$ Z'= Z_{\chi}cos\theta+Z_{\psi}sin\theta$$
Z$_{\eta}$ is obtained setting $\theta=arctg(-\sqrt{5/3})$. \par
The mass mixing coefficient (see appendix) between Z and Z' is given by\cite{Peswel,Hagi}: 
$$\gamma=2s_Wsin^2\beta(cos\theta \frac{1}{\sqrt{6}}-sin\theta
\sqrt{\frac{5}{18}})
+2s_Wcos^2\beta(cos\theta \frac{1}{\sqrt{6}}+sin\theta \sqrt{\frac{5}{18}})$$
where one assumes that the each Higgs boson doublet carries a vacuum expectation
 ($\beta=\pi/2$ corresponds to a pure H$_u$ referred as $\psi$u in the plot) and s$_W$=sin$\theta_W$. The mixing angle is then given by $\gamma M^2_Z/M^2_Z$'. 
Experts will notice that this description of mixing neglects other terms, like the kinetic mixing or some possible contributions from extra weak doublets needed to insure gauge coupling unification. This simplification is legitimate at the present stage of accuracy.
\subsection{LR}
For the LR scenario, the couplings depend on a single parameter, $\alpha_{LR}$.
For the lowest allowed value for this parameter, $\sqrt{2/3}$, the couplings are the same than for Z$_\chi$. The maximum value, which will be used in this paper, assumes LR symmetry for the couplings, g$_L$=g$_R$, and is given by:
$$ \alpha_{LR}=\sqrt{\frac{cos^2\theta_W}{sin^2\theta_W}-1}$$
For the mixing, assuming that symmetry breaking occurs through Higgs boson triplets,  one has\cite{Lan}:
$$\gamma=\sqrt{\frac{3cos2\theta_W}{5}}$$
\subsection{Extra dimensions}  
A scenario with an extra dimension at the TeV$^{-1}$ scale predicts Kaluza Klein (KK) recurrences for Z/$\gamma$. These recurrences couple to fermions with a strength $\sqrt{2}$ higher than the SM bosons, hence the higher mass sensitivity reached for these bosons.\par
These statements are of course model dependent since higher dimensions allow for non universal couplings 
if the different fermions live on separate branes (this allows in particular to suppress proton decay). 
 The measurement of Z-Z$_{KK}$ mixing provides critical information since, in these theories, mixing occurs only if the Higgs boson is confined to the 4D boundaries (see the appendix for more details). \par 
In the so called Universal Extra Dimension scenario, UED, where all SM fields propagate in extra dimensions, there is no direct vertex involving one non-zero KK mode. There is therefore no sensitivity 
through the interference method while the electroweak effect is 
suppressed\cite{appelq} for the first KK excitation of Z 
(see appendix). The measurable effect is due to t-b mass splitting which induces weak isospin violation in the KK spectrum. This theory is equivalent to the limit of the 'Little Higgs' scenario
in which one assumes that the gauge couplings go to zero. The current lower bound on the compactification scale, 300 GeV, comes from LEP/SLD PM, assuming a Higgs mass of 115 GeV. 
\par 
The string model of reference\cite{Iban} describes the three-family SM by considering
that there are intersecting stacks of D-branes 
corresponding to different gauge interactions: baryonic, left, right and leptonic. To this world are associated 4 Z'-type particles, one of them the SM Z, and three heavy with specific gauge charges, differing from the customary hypercharge coupling and therefore well recognised from, e.g., the 'Little Higgs' sector. These 4 Z' mix, implying a high sensitivity from LEP1. The W mass will be modified, as for E$_6$, by a positive contribution with a $\gamma$ which
depends on the Higgs boson sector of the model. In model A one finds that LEP1 results 
are very sensitive to the string scale, while this is not the case for model B.
One can then use LEP2 data and the fact that, in model B, Z' can have a 
mass of 1/10 the string scale to already achieve a limit of order 5 TeV on the string scale.
Effects on sin$^2\theta^{lept}_{eff}$ and $\Gamma_l$, in the string scenario, are more involved and will not be treated in this paper. \par 
\subsection{'Little Higgs'}
For this scenario three papers have been recently published\cite{csaki,hewettdec,than} allowing detailed calculations for LEP/SLD data. I have used the formulae provided by \cite{hewettdec}. These papers
 use a minimal version of the 'deconstruction' scheme\cite{amin}, the so-called SU(5)/SO(5) 'Littlest Higgs' in which one only generates the new particles needed to tame the quadratic divergences occurring in the Higgs sector: a triplet and a singlet of gauge bosons, a triplet of Higgs bosons and a vector-like quark t'.     
There are therefore two Z', a U(1) singlet, called hereafter B$_h$ and a triplet called Z$_h$. The effect on LEP results can be computed in terms of 3 parameters: the mass scale parameter f , the ratios x=g$_t$'/g' and y=g$_t$/g of the couplings in terms of the SM couplings. \par
t' gives no  significant contribution to PM unless the t-t' mixing angle takes a particular value (see the appendix). The main issue is therefore the 'fine-tuning' aspect since, as pointed out \cite{hewettdec}, LEP/SLD data, when combined with the Tevatron direct search, would correspond to a t' mass of order 10 TeV. Note already that the fine-tuning limit quoted in \cite{hewettdec} :
$$ m_{t'}<2\ \ TeV\ \ (m_H/200 GeV)^2 $$
would be very much relaxed with a heavier Higgs boson.  \par
In \cite{csaki} the possible contribution to PM of a Higgs boson triplet, an other necessary ingredient for the 'Little Higgs' scenario, is also discussed with the conclusion that, to get a consistent electroweak symmetry breaking, this contribution is necessarily much smaller than the Z' term. With LEP/SLD precisions, it is therefore legitimate to ignore this effect. \par  
Reference \cite{hewettdec} has used the Tevatron constraints to set a lower limit on the B$_h$ mass, 
while in the present paper I have combined LEP1/SLD PM and the hadronic cross-section from LEP2 to achieve the following limits: \par
 -B$_h$ heavier than 720 GeV assuming a 115 GeV SM Higgs boson \par
 -B$_h$ heavier than 810 GeV assuming a 500 GeV SM Higgs boson \par
It is interesting to notice that LEP1 and LEP2 data give complementary constraints. This is due to the fact that the limit from LEP2 improves by increasing x, the coupling of B$_h$ to fermions, while the Z-Z' mixing effect is minimum for x=1 and remains finite even at x=0. It turns out, as shown in figure 3, that the best 
LEP/SLD
solution corresponds to zero coupling to B$_h$, while there is almost no sensitivity for Z$_h$. \par 
These results demonstrate (figure 3) that with a
reasonable choice of x and y one can reach a good agreement with LEP/SLD data
without contradicting LEP2 and the Tevatron results. The preferred mass scale f$\sim$ 5 TeV is however severely violating the fine-tuning limit: 
$$ f < 1\ \ TeV\ \ (m_H/200 GeV)^2 $$
if one takes a 115 GeV Higgs boson. Reference \cite{than} claims that this problem could be cured by reducing
the $\rho$ contribution, e.g. by modifying the U(1) charge of the Higgs doublet. This would be at the price
of introducing new U(1) structures, beyond those embedded in SU(5). More recently\cite{wack}, a modification of the model has been proposed to have custodial SU(2) as 
an approximate symmetry. One can then reduce the impact on PM and f can be at the TeV scale. I will come back to the problem in section 4.2.
\par
\subsection{The BESS model}
I have chosen the so called degenerate BESS model which assumes the existence of two new triplets of gauge bosons degenerate in mass and with common couplings \cite{casal}. The effect on PM depends on one parameter as 
explained in the appendix. 
\subsection{Interpretations}
Interestingly we see that  in most cases there is Z-Z' mixing. Generally speaking one can say that there is mixing if the Z' couples to the SM Higgs boson. Even in the zero coupling limit, where there is no interference
effect, weak isospin violation still allows to observe a contribution to PM. This is the case for the 'Little Higgs'
and UED models. This is not the case for the APV measurement with Cesium due to an accidental cancellation which considerably reduces the sensitivity to the $\rho$ parameter\cite{altapv}.  \par 
If there is mixing, one expects two main effects: \par
- the Z mass will be modified inducing shifts on the 3 quantities, independent of the couplings \par
- the couplings are also modified via Z-Z' interference and this only affects sin$^2\theta^{lept}_{eff}$ and $\Gamma_l$. \par
In table 2 are given the main quantities needed to derive the influence of a Z' on the 3 LEP/SLD measurements which are used. 
\begin{table}[t]
\centering
\caption{Z' parameters for leptons}

\vskip 0.5 truecm
\begin{tabular}{|c|c|c|c|c|c|c|}
\hline Model &$\chi$ & $\psi$ u d & $\eta$ u d & $L-R$ &  B$_h$ & Z$_h$   \\  
\hline
\hline $\gamma$ & 0.39 & -0.51 0.51 & 0.64 -0.16 & 0.73  & &  \\
\hline q$_L$ & 0.61 & 0.26 & 0.17 & 0.33  & 0.5 & -0.5 \\
\hline q$_R$ & 0.20  & -0.26 & 0.33 & -0.43  & 1 & 0\\
\hline coupling/SM & s$_W$ & s$_W$ & s$_W$ & s$_W$ & xs$_W$ & yc$_W$ \\ 
\hline
\end{tabular}
\end{table}
 From figure 1,2 and 3, one already can draw a few interesting conclusions, even with the accuracies reached at LEP/SLD: \par
- there is much better discrimination by using 3 observables instead of 2, disfavouring for instance Z$_\chi$ ,Z$_{\psi u}$ and the BESS model. \par
- there is better agreement  with Z$_{LR}$ at 1.9 TeV, Z$_{\psi d}$ at 1.3 TeV, Z$_{\eta d}$ at 
0.8 TeV, a KK excitation at 3.4 TeV or an UED with a compactification scale at 0.4 TeV. \par
- there would be perfect discrimination between the various models with the FLC accuracy.\par 
In considering these conclusions, one should not forget that the b quark asymmetry result, which corresponds
to a heavy Higgs solution, has been removed. This explains, for instance, that some of these results seem to 
contradict previous conclusions. As an example, in \cite{casal}, the claim is that the BESS model can 
encompass a heavy Higgs, while figure 3 shows that the agreement is already marginally true for a 115 GeV 
Higgs mass for the leptonic width. This comes from the fact that without the b quark result, data prefer a 
40 GeV Higgs mass.  \par  
While LEP/SLD data are clearly insufficient to distinguish between the models,
they allow to exclude a light Z' with very high sensitivity. For instance, as shown in figure 1, a Z$_{\psi d}$ is already excluded at 1 TeV, a limit which compares favourably to the results of table 1, but which is more model dependent (e.g. one has to assume SM with a given Higgs boson mass). This is a well known feature but common wisdom tends to ignore that mixing is very likely in most models. In table 3 are given the mass limits using the 3 variables from LEP1/SLD assuming a 115 GeV Higgs boson mass.
\begin{table}[t]
\centering
\caption{Mass limits and preferred values in GeV from LEP/SLD} assuming a 115 GeV Higgs boson.
\vskip 0.5 truecm
\begin{tabular}{|c|c|c|c|c|c|c|}
\hline Model &$\chi$ & $\psi$ u d & $\eta$ u d & $L-R$ & B$_h$ & UED \\  
\hline
\hline Mass limit& 900 & 1500  950 & 1350 480 & 1200 & 720 & 300\\
\hline Best mass & - & \ \ - \ \ 1300 & \ \ \ - \ \ \ 800 & 1900 & 2900 & 430 \\
\hline
\end{tabular}
\end{table}
How do these results correlate with APV ? At the moment there is no significant deviation\cite{lanapv} but, if an effect is confirmed, it would favour Z$_{LR}$ or Z$_\eta$, since Z$_{\psi d}$ is purely axial and cannot be observed through APV. \par
\section{Can a Z' hide a 500 GeV Higgs boson ?}
Since the contribution of a Z' can compensate for a heavy Higgs boson, it is tempting to try pushing even further
the previous exercise and assume, for instance, a 500 GeV Higgs boson. 
\subsection{Standard schemes}
 The results for standard extended groups are shown in figures 4 and 5 which indicate that these scenarios miss the target. From figure 4 one could expect a better solution in between Z$_{\psi d}$ and Z$_{\psi u}$ but it turns out that this is not the case and that these 2 solutions are the closest to the data. The same is true for Z$_{\eta d}$ and Z$_{\eta u}$. The main rejection comes from $\Gamma_l$, dominated by the axial coupling. \par
A model independent approach would consist to allow the mixing, the left-handed
and right-handed couplings to vary independently. The result is that a TeV Z' could do the job, but with couplings which do not come out from
usual models: q$_R\sim$-3q$_L$ should be satisfied.    
This does not occur naturally, even relaxing some constraints (like g$_R$=g$_L$ in LR). \par
The BESS model, already marginally compatible with a 115 GeV Higgs, is clearly excluded at 500 GeV and was not represented in figure 6. \par 
 One can therefore state that LEP/SLD PM can exclude a 500 GeV Higgs boson even in the presence of a Z' of the standard types, which is an important constraint in view of LHC/FLC. \par
In noticeable contrast with this conclusion, one finds (see figure 6 in the zero coupling limit) that there is
excellent agreement with UED with a compactification scale of 200 GeV. If there is a heavy Higgs, this 
would mean that some of these particles could be seen at Tevatron and the whole spectrum thoroughly investigated at FLC.   
 \subsection{'Little Higgs'}
In this scheme, there is more flexibility (see the appendix) since one can vary independently the coupling, while in GUT groups the coupling constant to Z' is fixed. In practice this means that one can 
separate the effect on the $\rho$ parameter (called $\delta$ in the appendix)
which acts on M$_W$ and 
sin$^2\theta^{lept}_{eff}$  and the mixing which only acts on the later.  \par 
Figure 6 shows that  it is possible to select an x value compatible with a SM Higgs boson at \linebreak 500 GeV. Recalling the upper limit given for the t' mass and the mass scale f, one sees that there is no fine-tuning problem. \par
For what concerns the vector 
bosons, their masses are related to the scale f and to the couplings (see the appendix). 
\begin{table}[t]
\centering
\caption{Lower mass bounds in TeV within the  'Little Higgs' scenario allowing for a variable Higgs boson mass}
\vskip 0.3 true cm

\begin{tabular}{|c|c|c|c|c|c|}
\hline Masses  & m$_{B_h}$ &  m$_{Z_h}$ & m$_{t'}$& m$_{\phi}$ & f   \\
\hline
\hline LEP/SLD PM  & $\ge$ 2 & $\ge$ 3   & $\ge$ 9 & $\ge$ 7 & $\ge$ 3  \\ 
\hline
\end{tabular}
\end{table}
LEP/SLD data essentially constrain f, but one can vary substantially the parameter x and therefore the mass of B$_h$. The sensitivity to the coupling y, as shown in figure 3, is even smaller and therefore the mass of Z$_h$ is very uncertain. \par
Figure 7 shows the influence of the Higgs boson mass on f. There is a lower limit on the Higgs boson mass of order 300 GeV to satisfy the fine-tuning 
criterion. f has to be in a narrow interval to satisfy the LEP/SLD results (here also the same 3 observables have been used).
Table 4 summarises these informations. \par
Other precision measurements, like R$_b$, remain basically unaffected when m$_H$ and Z' parameters are varied. For what concerns A$^b_{FB}$, the 
problem stays the same since this quantity depends primarily on sin$^2\theta^{lept}_{eff}$. \par
It seems therefore that one could easily accommodate a heavy Higgs boson within the present scenario. This 
result is of course in strong contrast with the SUSY perturbative solution for which a light Higgs boson is mandatory and even in contrast with other types of Z' solutions.
\section{A Z' at FLC}
The origin of a Z' can be determined at FLC, as discussed in \cite{tesla}, by measuring the axial and vector couplings of the Z' to fermions at high energy. 
 The next step is to determine the Z-Z' mixing angle $\xi$, which gives information on the coupling of the Z' to the SM Higgs boson sector, and this can only be done by running at LEP energies.
\subsection{Standard schemes}
As can be seen in figure 8, FLC has a sensitivity which usually extends beyond the mass range of LHC. It is however fair to say that the separation between the various scenarios through precision measurements, using
cross sections and polarisation asymmetries, is unambiguous in a more restricted range.
While a Z$_{\chi}$, a Z$_{LR}$ or a KK excitation can certainly be identified by FLC with more than 5 s.d. in the whole LHC range, the discrimination between Z$_{\eta}$ or Z$_{\psi}$ would fall to about 3 s.d. for a \linebreak 5 TeV Z', assuming the precisions quoted in \cite{tesla}. 
  \par
\begin{table}[t]
\centering
\caption{Expected precisions on E$_6$ parameters at FLC for M$_{Z'}\sim$2 TeV}
\vskip 0.3 true cm
\begin{tabular}{|c|c|c|c|}
\hline
 Item &Z-Z' Mixing $\xi$ & $\chi-\psi$ mixing $\theta$ & up-down Higgs mixing cos2$\beta$  \\  
\hline
\hline FLC & $\sim$ \% &$\sim$ 0.1 rad  & $\sim$ 0.1 \\
\hline
\end{tabular}
\end{table}
 Figure 8 tells us that in most cases 
there will also be overlap between FLC and LHC for measuring the Z-Z' mixing angle $\xi$.  \par
 In the language of E$_6$, the couplings determine precisely $\theta$. Knowing $\theta$ and $\xi$, it is possible to extract the mixing angle $\beta$ which provides insights on the coupling of the Z' to the Higgs boson sector as explained previously. Table 5 summarises the expected accuracies. 
Note that they can be achieved if the Z' mass is given by LHC. \par 
For what concerns the LR model, FLC would also precisely determine the couplings and the mixing angle. Determining $\gamma$ is of great interest since this would allow to separate the symmetry breaking
schemes proposed in this model.\par
For the string model\cite{Iban}, one can expect a distinct behaviour of the axial and vector couplings since, as already mentioned, these couplings differ markedly from the hypercharge. 
Measuring the mixing angle seems essential in this case since it brings further informations on the Higgs boson sector and since it may well turn out that the lowest mass Z' will be the only signal from the string sector. \par 
The case of a Z/$\gamma$ KK recurrence is optimal for FLC which has a sensitivity well beyond LHC. As already pointed out, the mixing parameter gives a crucial information on the Higgs boson sector.\par
For the UED scheme\cite{appelq}, GigaZ measurements will allow to measure the compactification scale 
M$_{C}$ with a relative accuracy going like (M$_{C}$/3.3 TeV)$^2$, which covers the LHC domain of sensitivity.
\subsection{'Little Higgs'}
LEP/SLD data are not precise enough to give valuable constraints on B$_h$ and Z$_h$ masses. FLC informations will therefore play a critical role and one will have to use GigaZ data to assess the 'Little Higgs' interpretation. The Higgs 
boson mass will be known and, with the solution suggested by LEP/SLD data, the main information will come from GigaZ since the B$_h$ couplings to fermions 
are weak implying a small
interference effect at high FLC energy. \par
The precisions achievable on the coupling parameter x will be at the 
25 \% level for m$_{Bh}$=4 TeV, while the 
measurement of f, directly related to the W mass, will be very
precise. Therefore the mass of B$_h$ will also be predicted at the 25 \% level. \par
The sensitivity of GigaZ to Z$_h$ is reduced even if y takes large values
(see figure 3).  As discussed in the appendix, the triple gauge coupling measurements is sensitive to Z$_h$
for masses below 4 TeV. The best sensitivity can be reached using the 
interference method at high energy. With low x, the effect of B$_h$ is expected to be negligible at high energy. 
Assuming that f varies within the interval defined in figure 7, one finds, as indicated in table 6,  that 
the sensitivity goes up to 
5 TeV, which is largely sufficient to complement the LHC domain. \par  
\begin{table}[t]
\centering
\caption{Expected precisions on B$_h$ and $\phi$ masses in the 'Little 
Higgs' scenario. The prediction on m$_{\phi}$ assumes that the contribution
coming from v' can be neglected which, with the LEP/SLD solutions, corresponds to v'/v well below 1 \%. }
\vskip 0.3 true cm
\begin{tabular}{|c|c|c|c|c|}
\hline masses  & m$_{B_h}$ & m$_{\phi}$ & f & m$_{Z_h}$  \\
\hline
\hline $\Delta$M/M & $\sim$ (M$_{B_h}$/7.5 TeV$)^2$  & $\sim$ 0.01  & $\sim$ 0.01 &$\sim$ 
(M$_{Z_h}$/6 TeV)$^4$  \\  
\hline
\end{tabular}
\end{table}
\par

\section{Possible scenarios at  LHC}
The next question could be: can FLC results have an impact on LHC ? 
From present experience, we know that precision measurements from LEP/SLD, in conjunction with the top mass measurement from the Tevatron, have heavily influenced the program of FNAL which is now optimised for the search of a light Higgs boson. The interpretation of a Z' signal could have a similar impact on the strategy of LHC in terms of improving on the luminosity collected or in terms of data analysis. I will illustrate this idea
with two examples taken from the 'Little Higgs' and UED . \par
 
\subsection{The heavy Higgs scenario}  
Assume that the following situation occurs:  \par
'A Higgs boson, with mass heavier than 300 GeV, is seen at LHC and nothing else' \par 
One concludes that this result is inconsistent with LEP/SLD results and that there is new physics conspiring to create this situation. MSSM (with a very heavy SUSY spectrum) and most Z' schemes are inadequate. 
Assuming the 'Little Higgs' scenario, one can use the very precise GigaZ results to estimate the mass of B$_h$. If this mass is below 5 TeV, this gives some motivation to push to the limit the LHC luminosity to be able to discover this particle and confirm the hypothesis. \par
Z$_h$ will be out of reach at LHC unless y happens to be larger than x.  
As stated in previous section, interference measurements at high energy will at least be able to tell if 
Z$_h$ is within reach at LHC. \par    
FLC could therefore play a key role since it would not only 
provide an interpretation for the heavy Higgs boson mass but would also predict the 
sensitivity needed to observe additional particles at LHC and eventually  trigger upgrades
(e.g. increased luminosity) to improve their detection at LHC. More work is needed on this interesting 
scenario which illustrates convincingly the possible synergies between the 2 machines. \par

\subsection{The fake SUSY scenario}  
Assume first that the following situation occurs:  \par
'A light Higgs boson is seen at LHC and nothing else' \par 
This situation agrees with LEP/SLD and MSSM with very heavy SUSY particles. There is no signal at FLC 
but GigaZ
says that the direct and indirect Higgs mass measurements are incompatible. Within UED, GigaZ predicts whether LHC can observe the KK mass spectrum. If so one can try to improve the LHC search strategy accordingly. 
This case is very similar to the 'Little Higgs' scenario of previous section.\par
Assume alternatively that the following situation occurs:  \par
'A heavy Higgs boson is seen at LHC with SUSY-like particles ' \par
This is a confusing scenario since the Higgs mass is not compatible with MSSM and with the LEP/SLD PM.
The SUSY-like particles are in fact KK excitations of ordinary particles, relatively light (to compensate for the heavy Higgs in PM) which should be well measured and unambiguously interpreted at FLC. This would be an
ideal case of complementarity between LHC and LC.    
\section{Summary and future prospects}
A scenario with a Z' appears in many extensions of the SM being presently considered and should therefore retain our attention in anticipating physics at FLC/LHC. \par
There is so far no significant effect indicating the presence of a TeV Z', but there are suggestive deviations which could be reinforced, or weakened, with improved accuracy on the top mass at FNAL. It will also be of interest to follow the progress in the APV sector. \par
After LEP/SLD, there is no room to accommodate a heavy, of order 500 GeV or heavier, Higgs boson with the standard extended group scenarios with fixed couplings. \par
In contrast the 'Little Higgs' scenario is compatible with a heavy Higgs boson, a solution even favoured from the theoretical point of view since it avoids the 'fine-tuning' problem. This possibility offers a very challenging scenario for future colliders and illustrates convincingly a possible synergy between the 2 colliders.\par  
It appears that FLC can provide unique measurements on a Z', not only by running at its highest energy but also by improving the LEP/SLD precision measurements. This has been illustrated in this paper in case of the
'Little Higgs' scenario with low coupling of the lightest boson, indicated by the data, and the UED scheme.\par
Using these two examples, I have shown how FLC could provide a decisive clue in interpreting LHC findings and could even initiate a second round of activity at LHC, in a similar way LEP/SLD results on the Higgs boson sector have influenced the Tevatron program.\par 
These examples show that an FLC interpretation of LHC signals is likely to influence the strategy of LHC, in particular for the upgrades 
of the detectors, likely to occur if LHC runs at higher luminosities.
\vskip 2. cm

{\bf APPENDIX: useful formulae  }
\vskip 1. cm 

{\bf Extended groups}
\vskip 0.3 cm
Formulae used for a Z' within E$_6$ come from \cite{Peswel}. The Z mass shift and Z-Z' mixing are given by: 
$$\delta=\gamma^2\frac{m^2_Z}{m^2_{Z'}},\ \ \ \ \xi=\gamma\frac{m^2_Z}{m^2_{Z'}}$$ \par
The expected effects on the three observables I have considered are: 
$$\Delta m_W=57.6\delta (GeV),\ \ \ \Delta sin^2\theta^{eff}_W=-0.33\delta+0.23q_L\xi+0.26q_R\xi,
\ \ \ \Delta\Gamma_l=100\delta-173q_L\xi+150q_R\xi (MeV)$$ \par
with the literal expression of the first term given by:
$$ \frac{\Delta m^2_W}{m^2_W}=\frac{c^2_W}{c^2_W-s^2_W}\delta  $$ 
Note that sin$^2\theta^{lept}_{eff}$ depends mainly q$_L$+q$_R$, that is on the Z' vector coupling to leptons, while $\Gamma_l$ depends mainly on -q$_L$+q$_R$, that is the axial coupling.\par
$\xi$ can be derived by the general formula\cite{Hagi} which defines mixing:
$$ \gamma=-s_W\frac{\sum_i I_{i3}Q_i v^2_i}{\sum_i I^2_{i3}v^2_i}$$ \par 
where Q$_i$ and $I_{i3}$  are the  charge and isospin assignment of the Higgs boson doublets in the group under consideration. Vacuum expectations, like $v_u$ and $v_d$, come from the up and 
down Higgs boson doublets, but there could be additional from extra Higgs boson doublets, with $\sum_i v^2_i$=(246 GeV)$^2$. Recall that I have ignored the possibility of mixing through the kinetic terms\cite{Hagi}. \par  
The couplings for leptons are given by:
$$q_L=cos\theta\frac{3}{2\sqrt{6}}+sin\theta\frac{\sqrt{10}}{12},\ \ \ 
q_R=cos\theta\frac{1}{2\sqrt{6}}-sin\theta\frac{\sqrt{10}}{12}$$ \par
For the LR model, the electron couplings are given by:
$$q_L=\frac{1}{2\alpha_{LR}},\ \ \ 
q_R=\frac{1}{2\alpha_{LR}}-\frac{\alpha_{LR}}{2}$$ \par 
where $\alpha_{LR}$ is given by:
$${\alpha_{LR}}=\sqrt{(c_Wg_R)^2/(s_Wg_L)^2-1}$$ \par
with g$_L$=e/s$_W$.  
In this paper, I have assumed g$_R$=g$_L$, not necessarily true if parity-violation is broken at a scale above the LR symmetry scale. The mixing angle depends on the Higgs boson sector and I have assumed that LR symmetry breaking uses
Higgs triplets.
\vskip 0.3 cm
{\bf KK recurrences}
\vskip 0.3 cm
Formulae for KK recurrences come from \cite{Peswel}. A detailed treatment can be found in \cite{delgado}. Assuming that both Higgs doublets are confined to the 4D boundaries, one gets that:
$$ \frac{\Delta m^2_W}{m^2_W} =\frac{2\pi^2}{3} \frac{c^2_Ws^2_W}{c^2_W-s^2_W}
\frac{m^2_Z}{m^2_{Z'}} $$
This results in a very large parameter '$\gamma^2$', about 3 times larger than for LR, hence the large sensitivity given by LEP/SLC. Note that 
one cannot use anymore a simple description in terms of Z-Z' mixing since 
there is also a contribution to $\delta$ from W recurrences.
 \par
High energy interferences provide
an excellent sensitivity to the recurrence which couples $\sqrt{2}$ times more than an ordinary Z. \par
In the UED scheme with one extra dimension one has \cite{appelq}:
$$  \frac{\Delta m^2_W}{m^2_W} =-\frac{13\alpha c^2_W}{24\pi(c^2_W-s^2_W)}\frac{m^2_Z}{m^2_{Z'}} $$
which, by comparison to previous formulae, shows the suppression of the electroweak correction. There is
a measurable contribution, $\delta\sim 1.2\alpha (m_t/M_C)^2$, where $M_C$ is the compactification scale and $m_t$ the top mass, which is due to weak isospin violation induced by the top bottom mass difference. 
\vskip 0.3 cm
{\bf 'Little Higgs'}
\vskip 0.3 cm
For what concerns the 'Little Higgs' formalism, formulae are more complex since there are 2 Z' called B$_h$ and Z$_h$. Their masses are given in terms of the couplings and of a new mass scale f:
$$ m^2_{B_h}= \frac{g'^2}{10}\frac{(1+x^2)^2}{x^2}f^2, \ \ \  m^2_{Z_h}= \frac{g^2}{2}\frac{(1+y^2)^2}{y^2}f^2 $$
which shows why B$_h$ is usually lighter and plays the major role. \par
In this scheme the mixing parameter depends on the couplings and, with the notations given above, one has: 
$$\delta_G=\frac{s^2_W}{4}\frac{(x^2-1)^2}{x^2}\frac{m^2_Z}{m^2_{B_h}} \ \sim 
\ \frac{5}{8}\frac{v^2}{f^2} \ \ \ at \ \ \ low \ \ x \ \ \ and \ \ \ \gamma=-\frac{s_W}{2}\frac{x^2-1}{x}    $$
which shows that mixing vanishes when B$_h$ has the SM coupling (x=1). 
Also important is the fact that when the coupling constant goes to zero the mixing parameter tends to be very large for a given B$_h$ mass (expressed in terms of f, the result remains finite for x going to zero). \par   
The q$_{L,R}$ couplings to fermions are simply proportional to xI($^f_3$-Q$^f$) (hypercharge) for B$_h$ and 
to yI$^f_3$ for Z$_h$. At high energy one can in principle measure independently these couplings unless x and y are too small, in which case only the GigaZ measurements are available.\par
The mass of the t' quark depends on the mixing with t, $\theta_t$, and on the parameter f. For a small mixing angle one has:
$$m^2_{t'}=\frac{4m^2_t}{\epsilon(r-\epsilon)} \ \ \ \ where \ \ \ \ \
\epsilon=tan2\theta_t \ \ \ \ and \ \ \ \  r=v/f$$
The minimum value for the t' mass is reached when r=2$\epsilon$.
Unless r comes very close to $\epsilon$, the contribution to PM is usually small. \par
The scalar sector, that is the Higgs boson triplet needed in the 'Little Higgs' scenario, is discussed in \cite{csaki}. \par
In reference \cite{than}, an explicit formula is given for the mass of the triplet:
$$ m^2_\phi=\frac{4m^2_Hf^2}{v^2}\frac{1}{[1-32(v'f/v^2)^2]}   $$
where v' is the vacuum expectation of the triplet field. Note that I use convention from \cite{hewettdec} for f differs by a factor $\sqrt{2}$ from 
\cite{csaki}. 
 \par
This sector provides an additional contribution:
$$\delta_{\phi}=-4\frac{v'^2}{v^2} $$
which has an upper bound to keep $m^2_\phi$ positive such that
it is no more than 20 $\%$ than the B$_h$ contribution.
Applying the LEP/SLD constraints on f, one finds that the triplet part of 
$\delta$ is $\le$ 10$^{-3}$ and therefore cannot be a priori neglected with GigaZ accuracies. PM only measure
 the sum $\delta_G+\delta_{\phi}$ 
and give  no access to v'. \par
The triple gauge boson couplings also receive $v^2/f^2$ corrections\cite{than}. One can write:
$$2(c^2_W-s^2_W)(g^Z_1-1)=\delta-\frac{m^2_W}{m^2_{Z_h}}  $$
Knowing $\delta$ from the W mass measurement, one can deduce some information on the mass of Z$_h$. Using the solution suggested by LEP/SLD and the accuracies taken from \cite{tesla}, one finds that 
FLC gives a relative precision on m$_{Z_h}$ which goes like (m$_{Z_h}/5 TeV)^2$. \par  
\vskip 0.3 cm
{\bf The BESS model }
\vskip 0.3 cm
Formulae were taken from \cite{casal} and the effect on PM was computed according to \cite{altba}. They depend
on one parameter X:
$$ X=(\frac{g}{g_2})^2\frac{m^2_Z}{M^2} $$
where $g_2$ and M are the common gauge coupling and mass of new heavy gauge fields L and R. As an example, 
one easily derives that :
$$ \frac{\Delta m^2_W}{m^2_W}=X[\frac{2s^2_W}{c^2_W-s^2_W}+c^2_W
-\frac{c^4_W+s^4_W}{c^2_W-s^2_W}] \ \ \ 
\Delta sin^2\theta^{eff}_W=X\frac{s^2_W}{c^2_W-s^2_W}[c^4_W+s^4_W-1]$$
Numerically one gets:
$$ \frac{\Delta m^2_W}{m^2_W}=0.42X \ \ \ \Delta sin^2\theta^{eff}_W=-0.15X \ \ \ 
\frac{\Delta\Gamma_l}{\Gamma_l}=-0.76X  $$  
These relations translate into the BESS line shown in figure 3.
\vskip 0.5 true cm  

\begin{figure}
\epsfysize18cm
\epsfxsize18cm
\epsffile{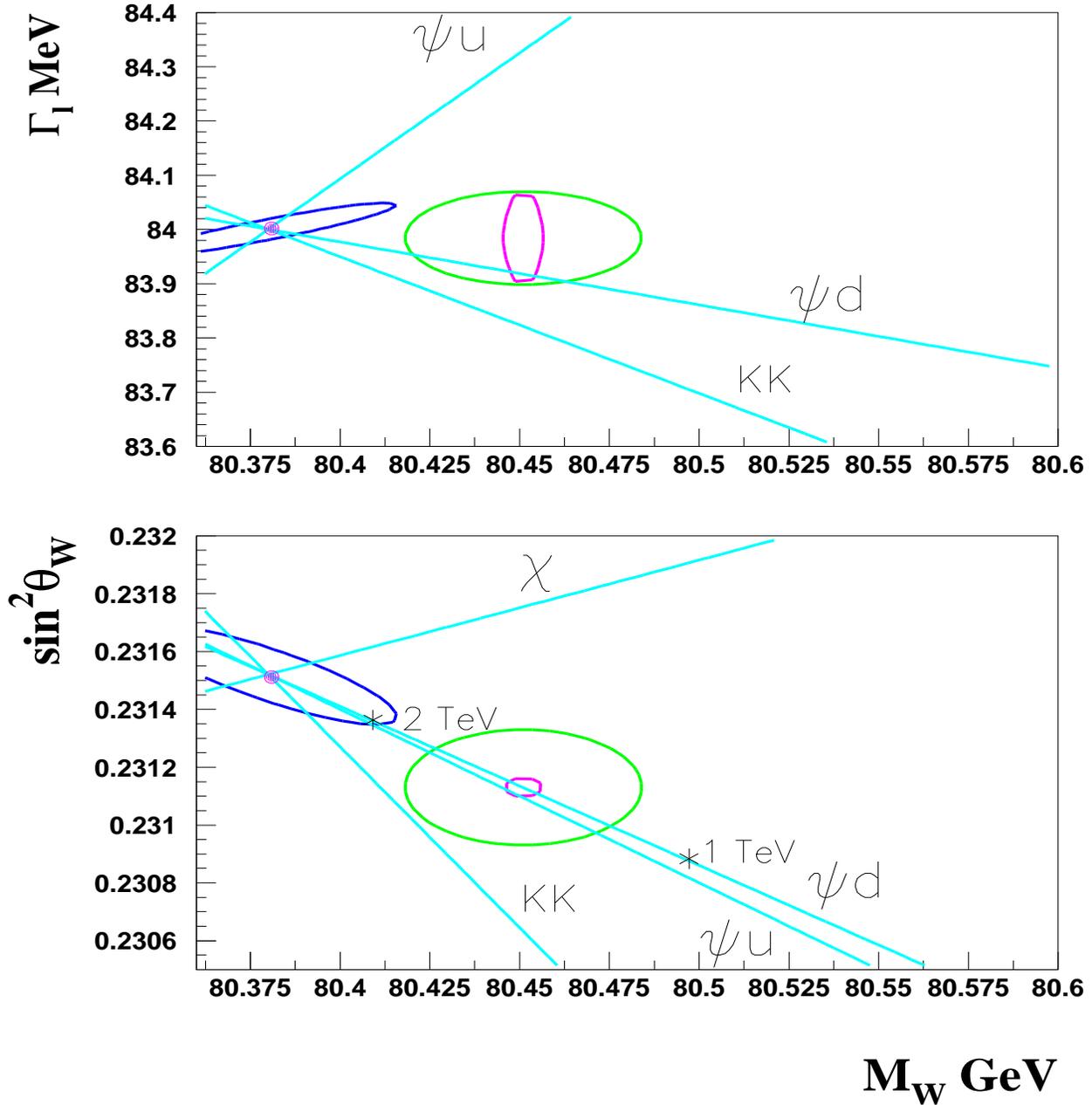}
\caption{On the top and bottom figures are indicated the SM prediction ellipse (on the left, dark blue) taking into account uncertainties coming from the top mass and from $\alpha(M_Z)$ and assuming a 115 GeV SM Higgs boson. The experimental measurements from LEP/SLD/Tevatron appear on the second ellipse (right, green) while inside these ellipses are the anticipated ellipses (purple) from GigaZ. Four models have been considered in the bottom part and only three on the top since the $\chi$ interpretation does not appear favoured. On the $\psi d$ line are indicated the points corresponding to Z' masses at 1 and 2 TeV.}
\end{figure}
\begin{figure}
\epsfysize18cm
\epsfxsize18cm
\epsffile{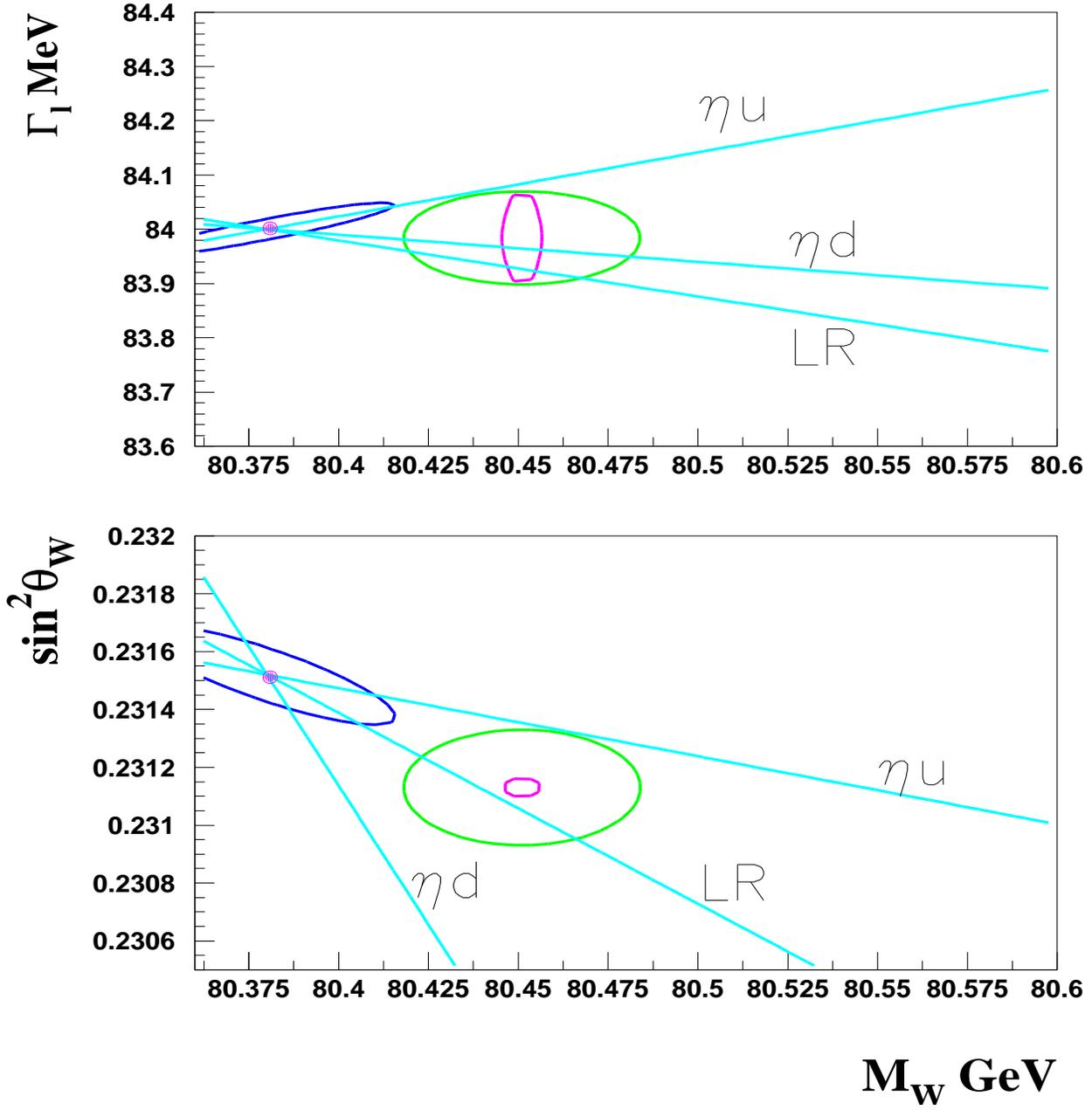}
\caption{Assuming, as in previous figure a 115 GeV Higgs boson, 3 Z' models defined in the text are tested against the LEP/SLD/Tevatron data.}
\end{figure}
\begin{figure}
\epsfysize18cm
\epsfxsize18cm
\epsffile{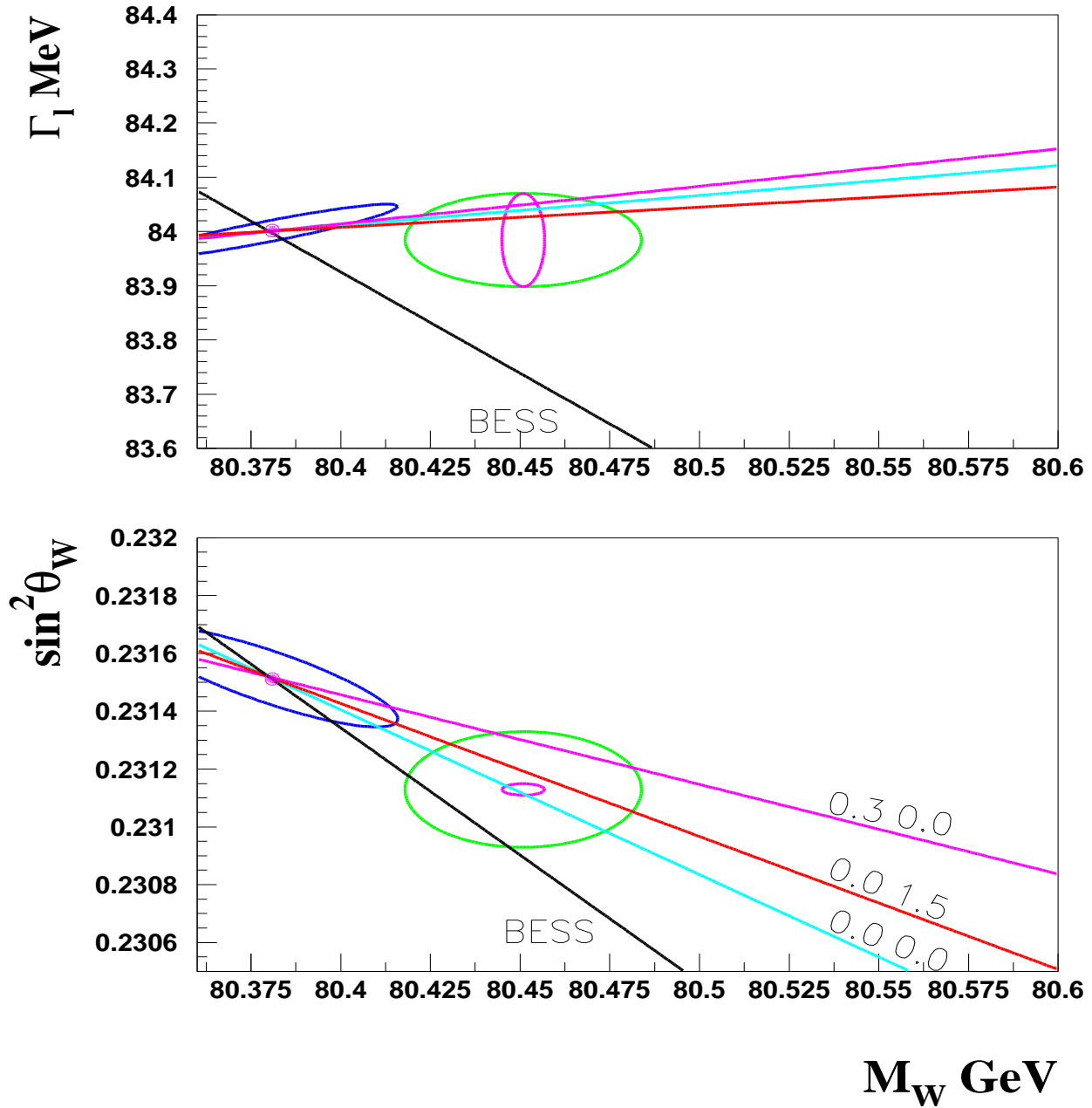}
\caption{'Little Higgs' assuming a 115 GeV SM Higgs boson. Three choices for x and y (indicated along the straight lines) are shown to illustrate the sensitivity to the B$_h$ and Z$_h$ couplings. Note that the zero coupling solution would correspond to the UED case. The BESS model is also represented.}
\end{figure} 
\begin{figure}
\epsfysize18cm
\epsfxsize18cm
\epsffile{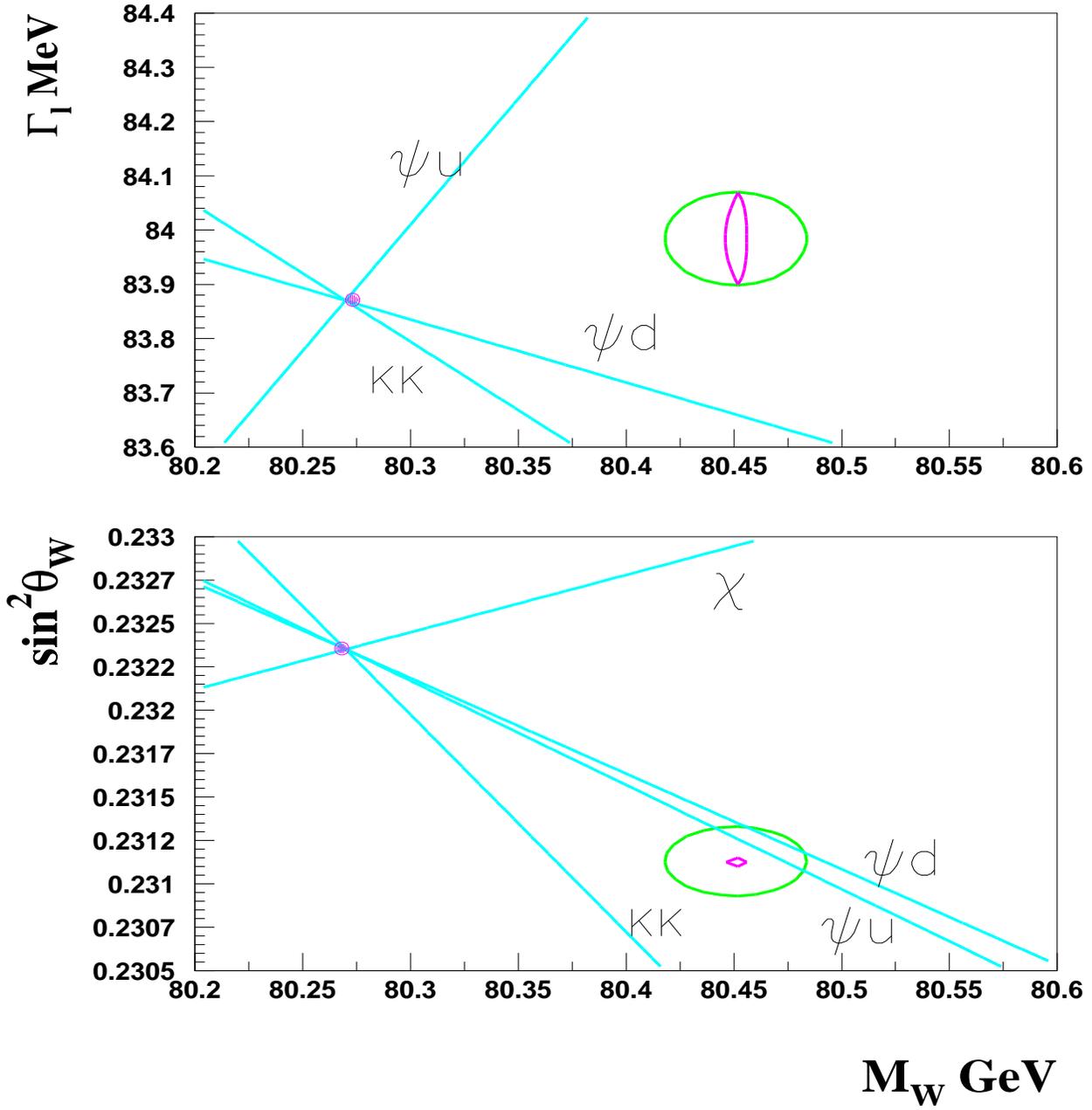}
\caption{The models used in figure 1 are tested assuming now a 500 GeV SM Higgs boson. The SM error ellipse plays a less critical role and has been removed from the plot.}
\end{figure}
\begin{figure}
\epsfysize18cm
\epsfxsize18cm
\epsffile{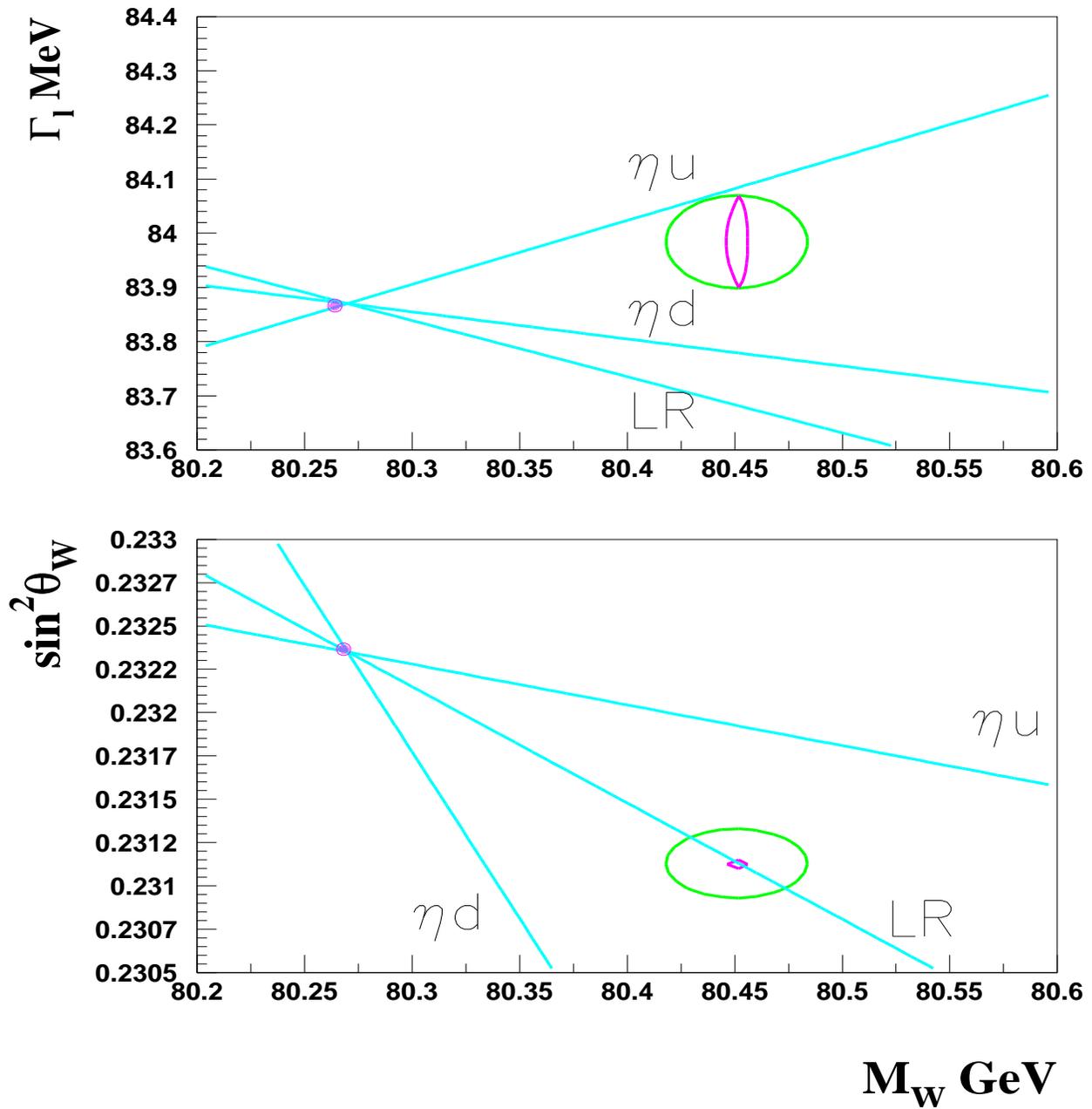}
\caption{ The models used in figure 2 are tested assuming now a 500 GeV SM Higgs boson.}
\end{figure}

\begin{figure}
\epsfysize18cm
\epsfxsize18cm
\epsffile{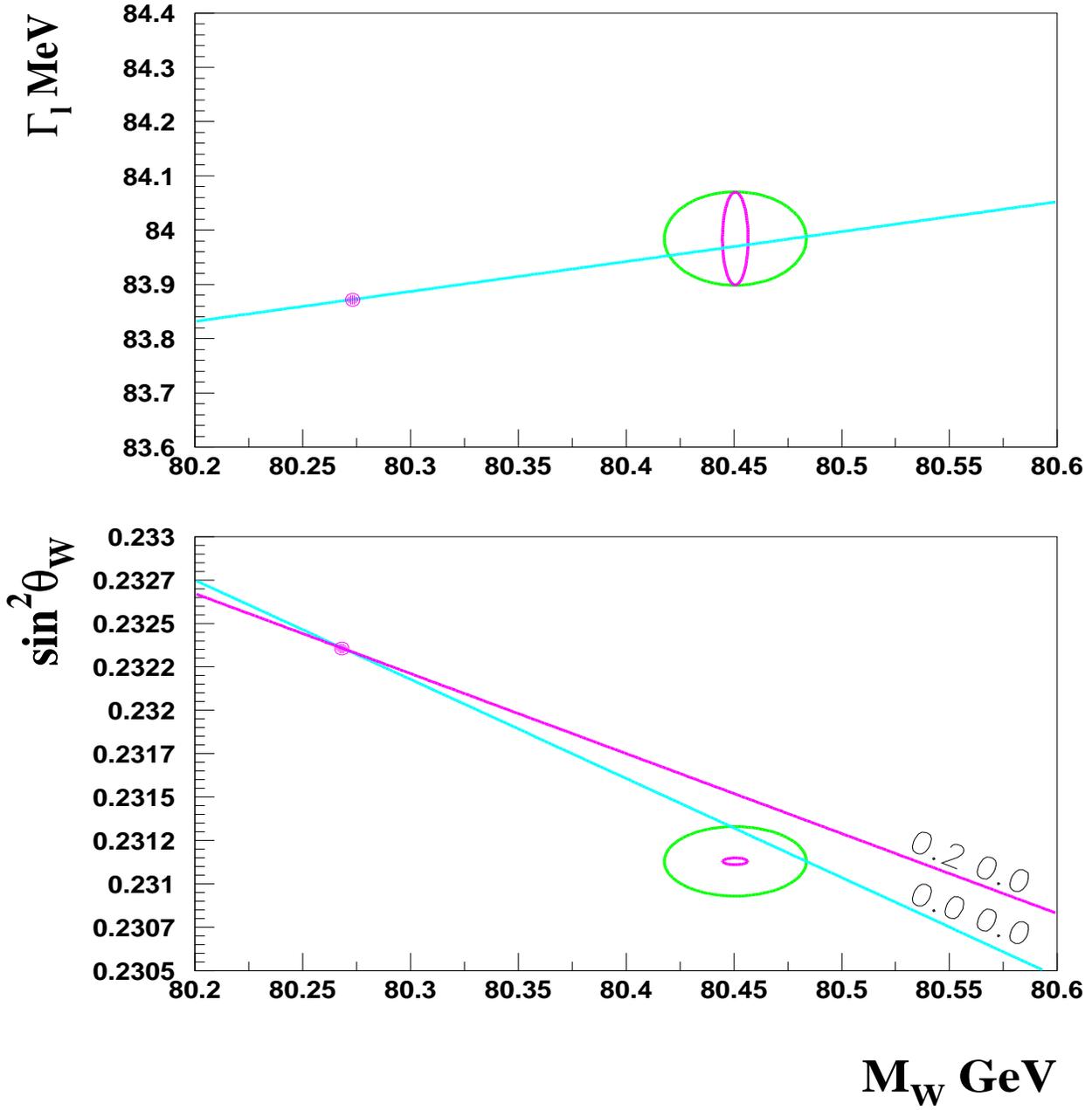}
\caption{'Little Higgs' with a 500 GeV SM Higgs boson. The best solution would correspond to \ \ \  3.5  TeV for the f scale. The zero coupling solution describes the UED case.}
\end{figure}
\begin{figure}
\epsfysize16cm
\epsfxsize16cm
\epsffile{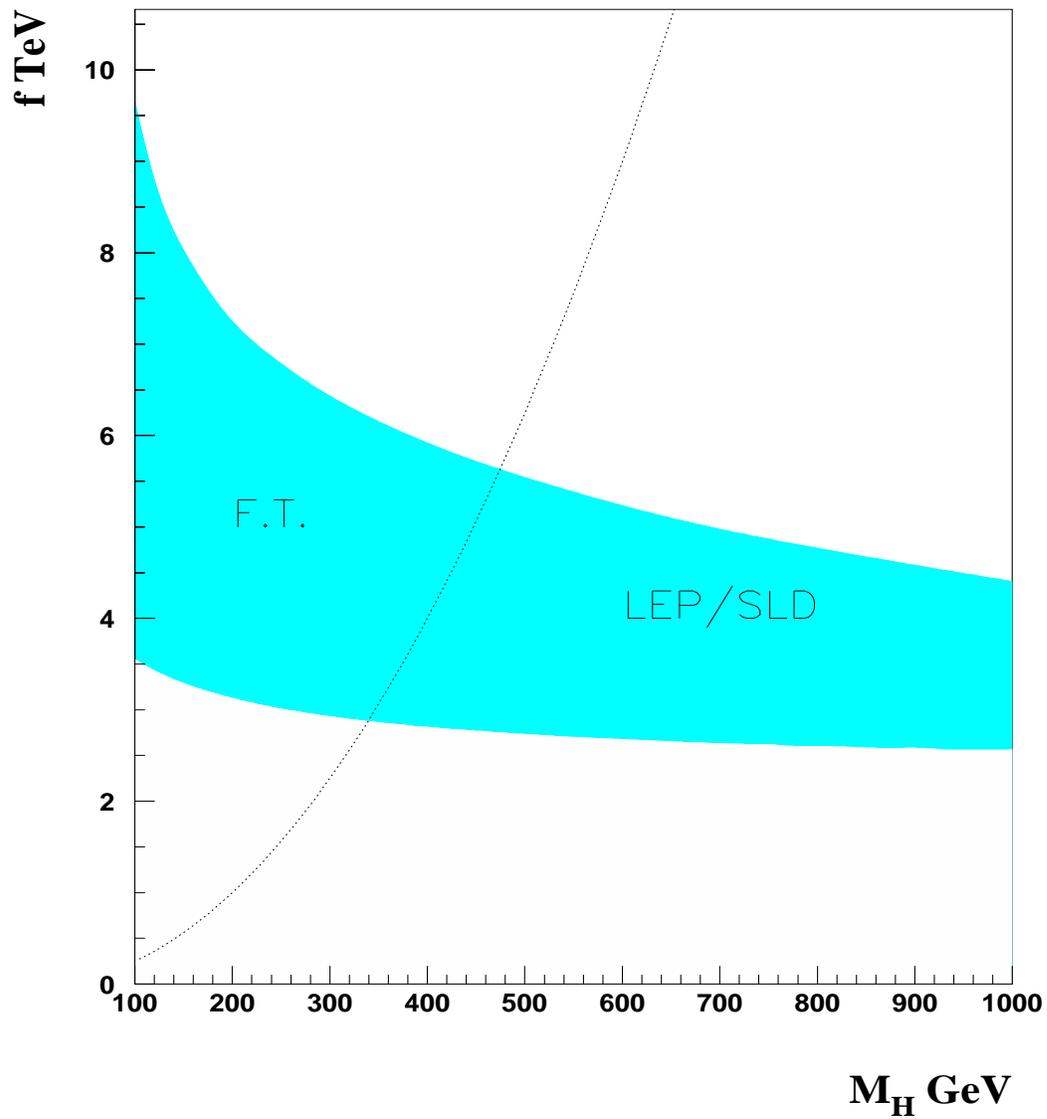}
\caption{Allowed regions for the mass scale parameter f in the 'Little Higgs'
scenario versus the Higgs mass in the low x approximation. The blue (dark) band
is allowed by the 3 measurements from LEP/SLD. The dotted curve indicates the fine-tuning limit and cuts away the LEP/SLD solution in the region marked F.T. The mass of B$_h$ is of the same order as f if one takes x=0.1.
It scales like 1/x. }
\end{figure}
\begin{figure}
\epsfysize16cm
\epsfxsize16cm
\epsffile{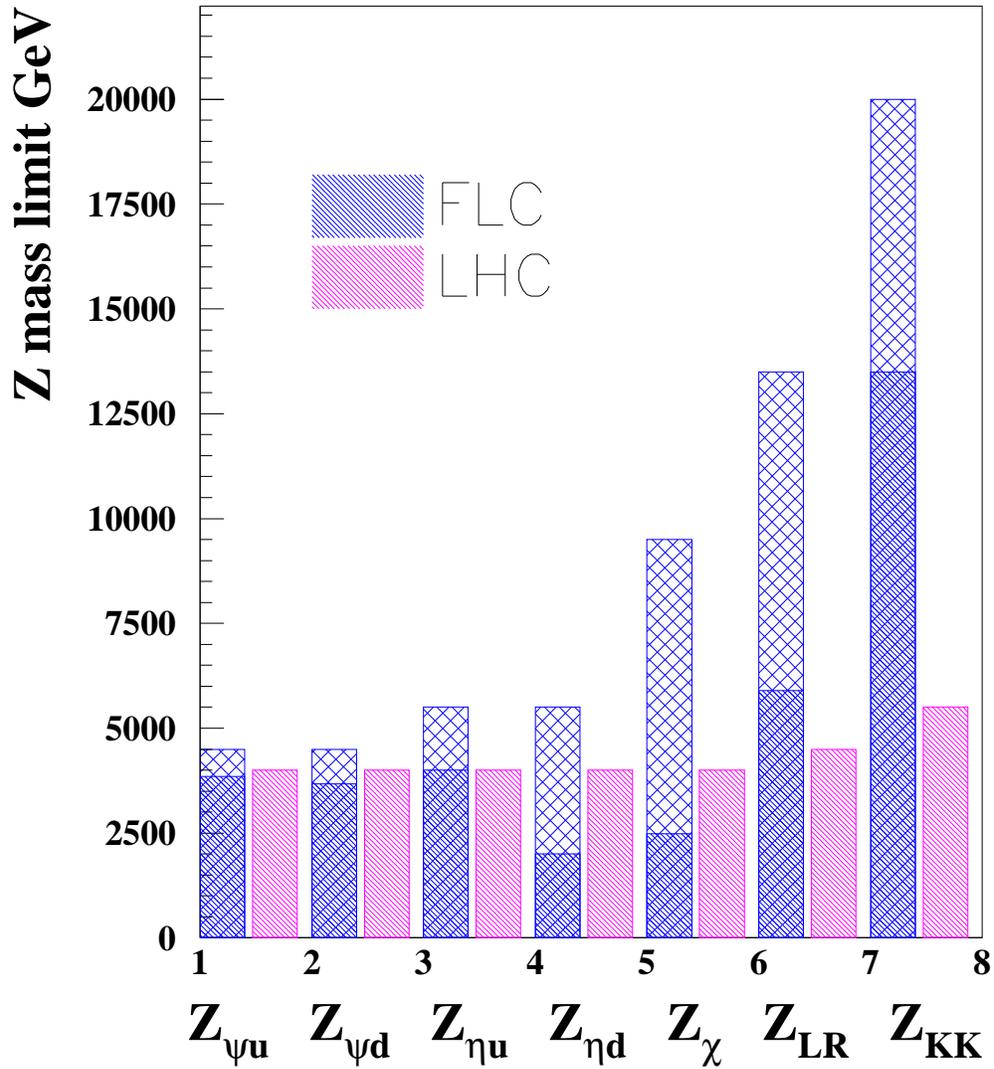}
\caption{Mass regions covered by LHC/FLC for a Z' in the various scenarios. For FLC (the left tower, in blue) the heavy hatched region is covered both by the mixing and the interference methods.}
\end{figure}

\end{document}